\documentclass[conference,a4paper]{IEEEtran}
\IEEEoverridecommandlockouts
\usepackage{cite}
\usepackage{amsmath,amssymb,amsfonts}
\usepackage{algorithmic}
\usepackage{graphicx}
\usepackage{textcomp}
\usepackage{xcolor}
\usepackage{hyperref}
\usepackage{comment}
\usepackage{subfig}
\usepackage{tabularx}
\usepackage{booktabs}
\usepackage[absolute,showboxes]{textpos}

\usepackage{pifont}

\usepackage{xspace}

\newcommand{\etal}{\textit{et al.}}
\newcommand{\eg}{\textit{e.g.,}~}

\newcommand{\one}{({\em i})\xspace}
\newcommand{\two}{({\em ii})\xspace}
\newcommand{\three}{({\em iii})\xspace}
\newcommand{\four}{({\em iv})\xspace}

\let\orgautoref\autoref
\renewcommand{\autoref}
{\def\sectionautorefname{Section}%
\def\subsectionautorefname{Section}%
\def\subsubsectionautorefname{Section}%
\orgautoref}

\makeatletter
\renewcommand{\paragraph}[1]{\vspace*{0.03in}\noindent{\bf #1.}\hspace{0.25ex \@plus1ex \@minus.2ex}}
\newcommand{\paragraphS}[1]{\vspace*{0.03in}\noindent{\bf #1}\hspace{0.25ex \@plus1ex \@minus.2ex}}
\makeatother

\def\BibTeX{{\rm B\kern-.05em{\sc i\kern-.025em b}\kern-.08em
    T\kern-.1667em\lower.7ex\hbox{E}\kern-.125emX}}
\hypersetup{nolinks=true}
\begin{document}

\title{WIP: Exploring DSME MAC for LoRa --\\A System Integration and First Evaluation}

\author{\IEEEauthorblockN{Jos{\'e} {\'A}lamos}
\IEEEauthorblockA{\textit{HAW Hamburg \& FU Berlin} \\
\small{jose.alamos@haw-hamburg.de}}
\and
\IEEEauthorblockN{Peter Kietzmann}
\IEEEauthorblockA{\textit{HAW Hamburg} \\
\small{peter.kietzmann@haw-hamburg.de}}
\and
\IEEEauthorblockN{Thomas C. Schmidt}
\IEEEauthorblockA{\textit{HAW Hamburg} \\
\small{t.schmidt@haw-hamburg.de}}
\and
\IEEEauthorblockN{Matthias W{\"a}hlisch}
\IEEEauthorblockA{\textit{Freie Universit{\"a}t Berlin} \\
\small{m.waehlisch@fu-berlin.de}}
}

\maketitle

\setlength{\TPHorizModule}{\textwidth}
\setlength{\TPVertModule}{\paperheight}
\TPMargin{5pt}
\begin{textblock}{1}(.1,0.01)
\noindent
\footnotesize
If you cite this paper, please use the WoWMoM reference:
J. Alamos, P. Kietzmann, T. C. Schmidt, M. W{\"a}hlisch.\\
WIP: Exploring DSME MAC for LoRa – A System Integration and First Evaluation.\\
\emph{Proc. of the 23rd IEEE International Symposium on a World of Wireless, Mobile and Multimedia Networks}, IEEE, 2022.
\end{textblock}

\begin{abstract}
    LoRa is a popular wireless technology that enables low-throughput (bytes)
    long-range communication (km) at low energy consumption (mW). Its transmission, 
    though, is on one side prone to interference during long on-air times, and
    on the other side subject to duty cycle restrictions. LoRaWAN defines a MAC and a 
    vertical stack on top of LoRa. LoRaWAN circumvents the above limitations by imposing 
    a centralized network architecture, which heavily reduces
    downlink capacity and prevents peer-to-peer communication. This makes it
    unusable for many deployments. The Deterministic and Synchronous Multichannel
    Extension (DSME) of IEEE 802.15.4e benefits of time-slotted communication and
    peer-to-peer communication and has the potential to overcome LoRaWAN
    limitations.
    In this work, we implement DSME on top of LoRa in the open source IoT OS RIOT 
    and open the field for first evaluation experiments on real hardware.
    Initial results indicate that DSME-LoRa not only enables reliable peer-to-peer communication for
    constrained IoT devices, but also scales with an increasing number of nodes.
\end{abstract}

\begin{IEEEkeywords}
Wireless, 802.15.4e, IoT networking
\end{IEEEkeywords}

\section{Introduction}\label{sec:intro}

LoRa is a popular wireless technology for the Internet of Things (IoT) that achieves long range transmissions (km) at minimal power consumption (mW).
The narrowband chirp spread spectrum modulation is robust against
interference and doppler effect.
LoRa operates in unlicensed subGHz spectrums, which are subject to
regional band regulations.
LoRaWAN is a cloud-based network architecture for LoRa that organizes 
all communication between constrained Endnodes (ENs) and user applications.
LoRaWAN consists of three components:
Application Servers (ASs) provide an interface for business logic implementation;
A centralized Network Server (NS) coordinates communication including the PHY configuration, media access, and routes between ASs and ENs;
Gateways (GWs) act as a LoRaWAN backbone and mediate packets between ENs and the NS.
Three constrains are worth stressing:
First, downlink packets are heavily regulated. Regional band restrictions limit the number
of downlink packets per GW.
Hence, high downlink loads lead to unpredictable and long latencies as well as packet  loss. This makes LoRaWAN impractical for many applications.
Second, the centralized architecture of LoRaWAN challenges data sharing between users, between ENs, and complicates the development of distributed applications across the Internet.
Higher-layer protocols (IP, CoAP) run inefficiently on top of LoRaWAN networks.
Third, LoRaWAN requires a permanent infrastructure backhaul. Intermittent GW connectivity prevents data forwarding between the LoRaWAN network and ENs. Since peer to peer communication is impossible between ENs, unreachable GWs prevent communication.

We argue for replacing the LoRa MAC to overcome LoRaWAN limitations.
Zorbas~\etal~\cite{zf-tlndc-21} summarize the potentials of time-slotted MAC layers for LoRa and provide a literature survey.
They propose TS-LoRa~\cite{zakp-ttlii-20} as a LoRaWAN alternative. It adds a time-slot extension to LoRaWAN and introduces group ACKs to save downlink, but the solution still requires a permanent infrastructure backhaul.
Haubro~\etal~\cite{hoof-tlrri-20} present an approach for IEEE 802.15.4e TSCH mode (Time Slotted Channel Hopping) over LoRa which suggests performance potentials.
We argue that IEEE 802.15.4e DSME (Deterministic and Synchronous Multichannel Extension) is better suitable for LoRa radios.
In contrast to TSCH, DSME provides both contention-access transmission (using CSMA-CA) and contention-free transmission (time- and frequency multiplex).
Three built-in features of DSME make it an appealing candidate for LoRa.
First, DSME creates slot schedules natively in a decentralized manner, which reduces management overhead and enables mesh and multi-hop topologies out of the box.
Second, the multisuperframe structure of DSME outperforms TSCH in
throughput and delay for high transmission duty cycles and large
networks~\cite{vbpa-infid-17}.
Third, DSME supports native group ACKs, which reduces downlink load.
In our previous work~\cite{aksw-dfml-21}, we proposed a DSME-LoRa mapping scheme as well as an information centric networking adaption~\cite{kaksw-liiel-22} that base on simulation results for time-slotted transmissions. In this paper, 
 we go a step further with the implementation of DSME-LoRa on real hardware.

The contributions of this paper are the following.
We introduce the necessary background. (\S~\ref{sec:dsme}).
We present our DSME-LoRa integration into the RIOT~\cite{bghkl-rosos-18} network stack (\S~\ref{sec:environ}) and perform first real-world measurements on constrained IoT nodes in the open access FIT IoT-LAB testbed. Next, we compare our practical measurements to simulation results (\S~\ref{sec:eval}) that we conduct on our previous work~\cite{aksw-dfml-21} and draw a first conclusion about the practicability of DSME-LoRa (\S~\ref{sec:conclusion}).

\section{Problem Statement and Challenges}\label{sec:dsme}

\begin{figure}[!h]
    \includegraphics[width=\columnwidth]{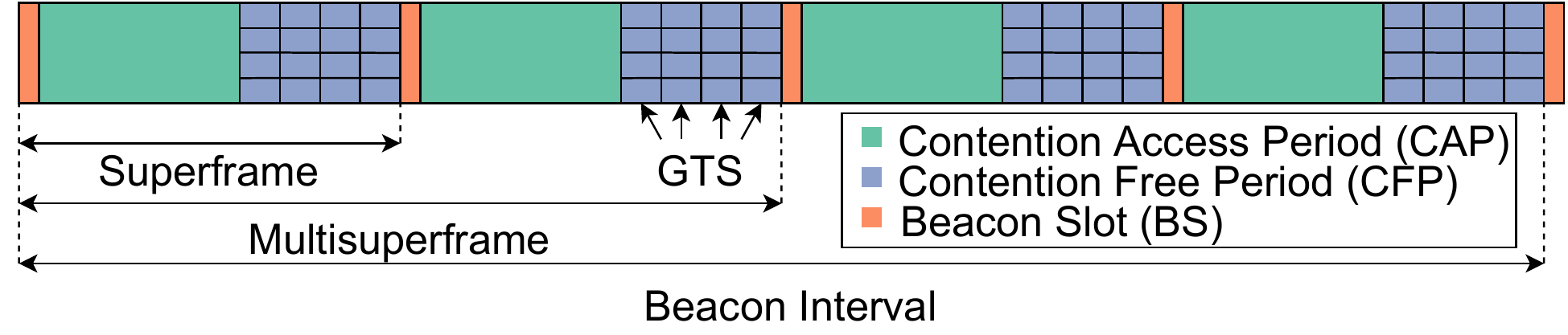}
    \caption{Example of a DSME multisuperframe structure consisting of two superframes per multisuperframe and a beacon interval of two multisuperframes.\vspace{-2mm}}
    \label{fig:dsme_fundamentals}
\end{figure}


DSME defines a deterministic and synchronous multichannel that allows for
coordinated communication.
Coordination is based on superframes, which consist of three
parts: Beacon slot (BS), contention access period (CAP), and contention free
period (CFP).
A series of superframes compose a multisuperframe, see Figure~\ref{fig:dsme_fundamentals}.
Data transmission occurs during the CAP using CSMA-CA on a single
common channel, or during the CFP using a guaranteed time slot (GTS). 
DSME devices synchronize to their neighbours through (enhanced) beacons
received during the beacon slots.

The integration of DSME on top of LoRa on real hardware imposes a series of
challenges that are not present on a simulation environment.
\one Longer on-air times of LoRa require a longer superframe duration, which
increments the beacon interval time.  Thereby, devices are prone to
de-synchronization due to clock drift between neighbours.
\two LoRa transceivers do not add RX timestamps to received frames, which DSME
requires to calculate the time difference to a neighbour and perform
synchronization.
\three In case of time critical operations, DSME accesses a transceiver based on
Interrupt Service Routine~(ISR).
On the one hand, this limits the responsiveness of real time operating
systems, and, on the other hand, faces concurrent access to the
hardware bus~(SPI).
\four Common IoT LoRa hardware is constrained in terms of memory, typically
around 100~kB of ROM and 10~kB of RAM, which is enough for a LoRaWAN stack.
In contrast to LoRaWAN, DSME requires more memory due to the complexity of the MAC.

\section{Implementation}\label{sec:environ}

\begin{figure}
\center
\includegraphics[scale=0.7]{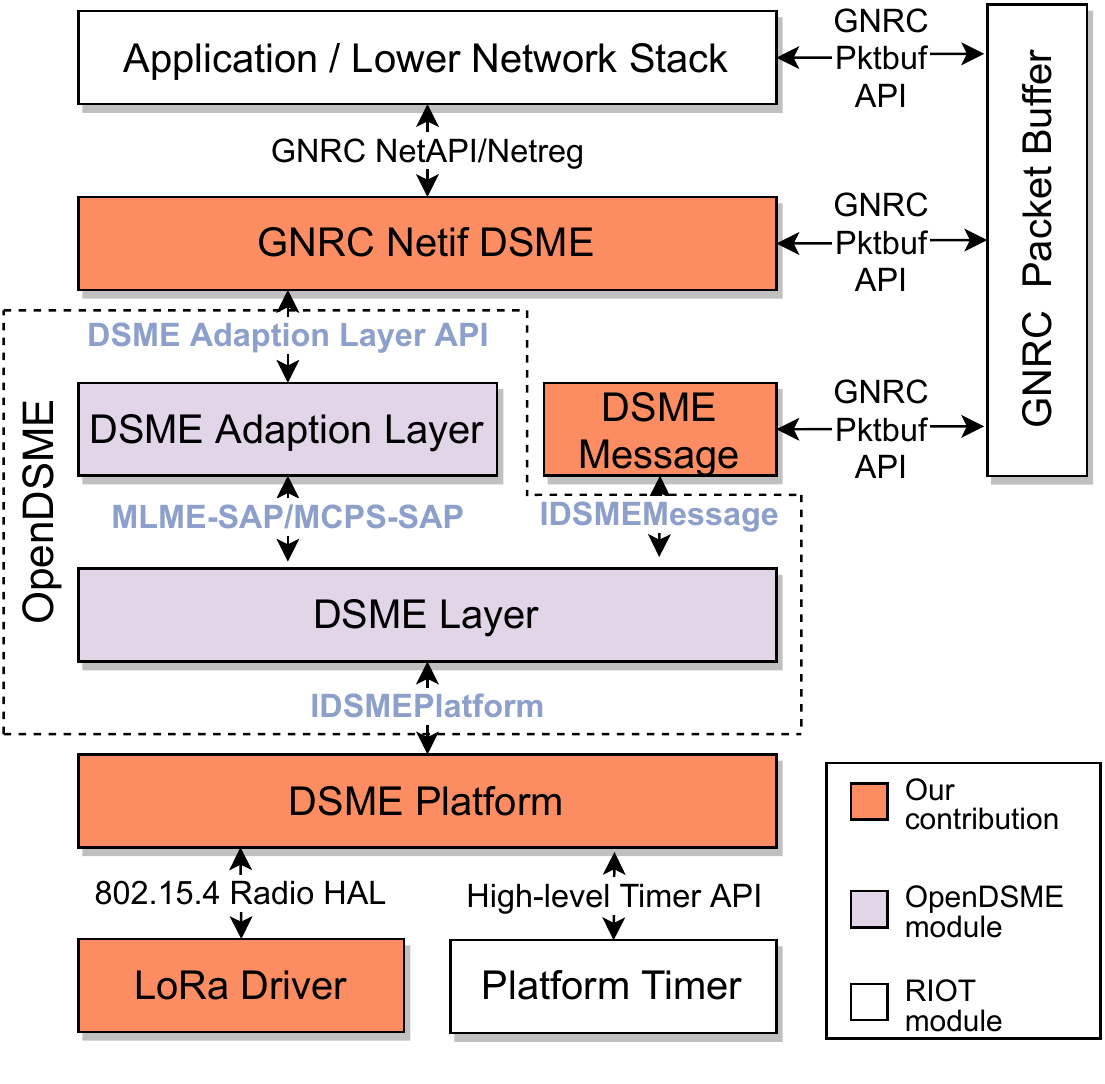}
	\caption{DSME-LoRa integration into the networking subsystem of RIOT.\vspace{-2mm}}
\label{fig:dsmelora_integration}
\end{figure}

We integrate the \textit{openDSME}~\cite{kkt-rwmnd-18} MAC implementation to the networking subsystem of RIOT (namely GNRC). GNRC provides a generic inter-module messaging interface (GNRC NetAPI), a packet dispatch registry (GNRC Netreg) and a centralized packet buffer (GNRC Pktbuf). RIOT provides a hardware abstraction layer for standard 802.15.4 radios, as well as a timer abstraction layer.

\subsection{DSME-LoRa Integration}\label{sec:integration}
\autoref{fig:dsmelora_integration} presents an overview of the software modules and exposes existing RIOT modules and APIs (white), \textit{openDSME} modules and APIs (purple), and our contributions (orange).

\textbf{GNRC Netif DSME}  implements the DSME network interface.
This enables applications or an IP stack
to communicate via the GNRC messaging interface  NetAPI and the packet dispatcher
registry GNRC Netreg.
Our network interface implementation utilizes the DSME Adaption Layer internally, which is a convenience API provided by \textit{openDSME}, to access the DSME Layer that implements the MAC logic. 

\textbf{DSME Message} implements the \textit{openDSME} packet interface (IDSMEMessage) that allocates memory in the GNRC Pktbuf, which facilitates a transparent integration of DSME-LoRa with the networking subsystem.
GNRC Pktbuf provides three promising features:
\one centralized storage that prevents data duplication.
\two A scattered packet representation which allows to append chunks of bytes in a packet, without memory reallocation.
\three It supports allocation with \texttt{malloc} instead of static memory allocation, to operate on the same memory pool as \textit{openDSME} which bases on heap.

\textbf{DSME Platform} implements the DSME platform interface (IDSMEPlatform) which acts as a hardware abstraction layer. It includes two parts.
\one The implementation of transceiver access routines on top of the
802.15.4 Radio HAL in RIOT.
We reconfigure \textit{openDSME} to adjust to LoRa symbol times, based on our proposed PHY configuration~\cite{aksw-dfml-21}.
We use the \textit{ValidHeader} IRQ of the transceiver to tag the LoRa PHY
header and calculate the RX timestamp of the LoRa frame.
\two The access to timer functionalities of the operating system. Thereby, we configure the high-level timer to use a precision real-time timer peripheral, to reduce de-synchronization due to clock drifts.
We leverage the processing of radio and timer events to the RIOT scheduler, to ensure DSME does not
access the transceiver during ISR.

\textbf{LoRa Driver} implements a 802.15.4-type device driver for the LoRa
transceiver. We map the LoRa PHY configuration proposed in~\cite{aksw-dfml-21} to 16
LoRa channels, and CCA to the channel activity detection feature (CAD) of the radio.

\subsection{Experiment Deployment}\label{sec:exp_environment}
We run DSME-LoRa on the Saclay site of the FIT IoT-LAB testbed which provides 20 x B-L072Z-LRWAN1 nodes that base on an ARM Cortex-M0+ CPU and contain a SX1276 LoRa transceiver.
The platform provides 192\,kB of ROM and 20\,kB of RAM.
To meet these constrains with our implementation, we
\one limit the number of superframes per multisuperframe to one. This reduces the number of available GTS, hence, decreases RAM requirements;
\two configure GNRC Netreg to use callbacks instead of IPC, which saves one additional receive thread.

We deploy a sensor-actuator network that consists of one coordinator node, 5/10/15 sensors that transmit 16\,Byte payloads to three actuator nodes, at an exponential packet rate. Hence, our network incorporates 19 LoRa devices at max.
Thereby, we analyze data transmission in CAP and CFP, and we vary the number of sensors as well as the average transmission rate. For the CFP case, a static resource allocation scheme allocates transmission cells during bootstrap, to prevent slot negotiation during experiment run. The cell assignment adds one unique cell to each sensor-actuator pair, in a multisuperframe.

We compare our real-world measurements to simulation results that we conduct in OMNeT++ /INET, based on the environment that was presented in~\cite{aksw-dfml-21}.


\section{Evaluation}\label{sec:eval}

\begin{table}
    \begin{center}
    \begin{tabular}{l r r r r}
        \toprule
        &\multicolumn{2}{c}{\textbf{ROM}}&\multicolumn{2}{c}{\textbf{RAM}}\\
        \cmidrule(l){2-3}
        \cmidrule(l){4-5}
        \textbf{Component} & {[kB]} & Prop. {[\%]} & {[kB]} & Prop. {[\%]} \\
        \midrule
        Application & 1.07 & 0.99 & 2.17 & 18.39 \\
        \textit{openDSME} & 64,59 &59.77 & 7.67 &65.13\\
        OS (incl. GNRC) & 39.17 &36.24 & 1.94 &16.48 \\
        LoRa Driver & 3.25 &3.00 & 0.00  &0.00 \\
        \midrule
        \textbf{Total} & \textbf{108.06} & \textbf{100.00} & \textbf{11.78} & \textbf{100.00} \\
        \bottomrule
    \end{tabular}
    \caption{ROM and RAM requirements for DSME-LoRa integration in RIOT, measured on a B-L072z-LRWAN1 platform. Prop. [\%] depicts the proportion of the total image size.}
    \label{tbl:firmware_size}
    \end{center}
\end{table}

\subsection{Memory Requirements}
\autoref{tbl:firmware_size} shows the firmware size, separated into ROM (text + data segment) and RAM (bss + data segment), for a
DSME-LoRa firmware image that we compile for our test platform.
Thereby, we group the memory consumption into four components:
The application utilizes $\approx$ 2\,kB of RAM for thread stack allocation. We utilize default values here, which gives enough space to operate standard IoT applications using sensors and actuators.
\textit{openDSME} contains the core MAC implementation, as well as the DSME Adaption Layer, DSME Platform, and DSME Message (compare~\autoref{fig:dsmelora_integration}). This contributes the biggest proportion of 60--65\% to the total firmware size.
It is noteworthy, however, that the implementation maintains a variety of data structures that enable a direct use in the INET simulation framework. The embedded integration would benefit from further optimization.
Operating system (OS) includes the kernel, scheduler, drivers, and OS utilities (\eg a shell), as well as the dependencies of the networking subsystem GNRC. Hence, we account our GNRC Netif DSME network interface to that group. In total, this requires $\approx$ 40\,kB of ROM, and $\approx$ 2\,kB of RAM, which is in line with former analyses~\cite{bghkl-rosos-18}.
Finally, the integration of our LoRa transceiver driver below the 802.15.4 Radio HAL operates frugal and only requires $\approx$ 3 \,kB of ROM.

The firmware requires a total of 108.06\,kB in ROM and 11.78\,kB in RAM which fits our target platform.
The `unused' RAM of 8.22\,kB is utilized for dynamic runtime memory allocation (heap).
Three operations of \textit{openDSME} make use of the heap.
\one allocation of a packet (92\,Bytes plus size of the MAC data frame without frame checksum), 
\two allocation of a GTS slot (44\,Bytes), and
\three allocation of a neighbour queue entry (124\,Bytes).
A neighbour entry is instantiated on every GTS association with a new neighbor node.
As an example, a device that allocates 6 GTS with 3 different neighbours schedule 5 frames (each with a size 25\,Bytes).
This requires $6 \cdot 44 + 3 \cdot 124 = 636$\,Bytes of for slot allocation, and $5 \cdot (92 + 25) = 585$ Bytes
for packet allocation, which are allocated on the heap using \texttt{malloc}.

\subsection{Data Transmission Performance}

\begin{figure*}
    \centering
    \subfloat[CAP, TX interval=20s]{\includegraphics[width=.25\textwidth]{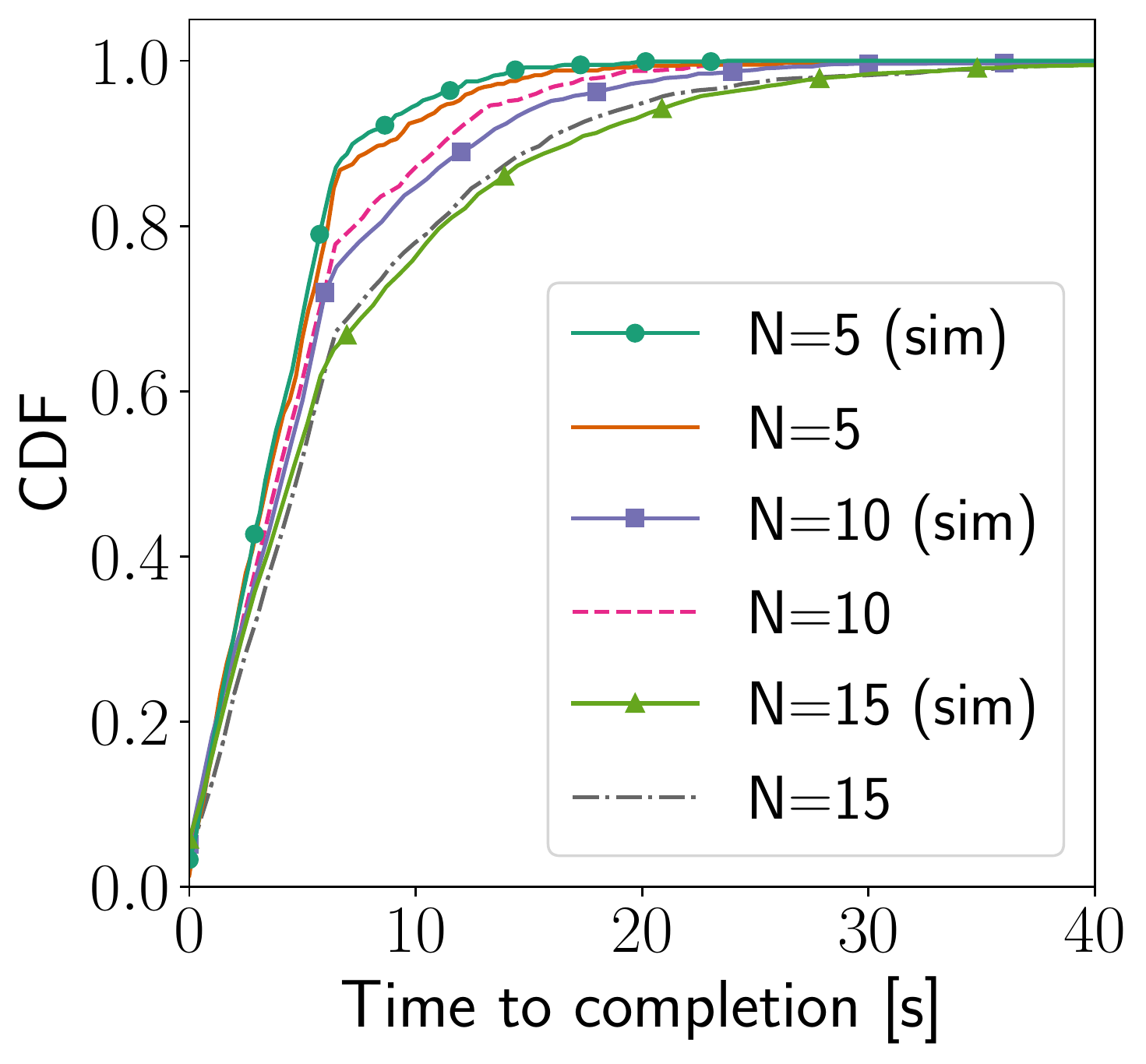}\label{fig:cap_20}}
    \subfloat[CFP, TX interval=20s]{\includegraphics[width=.25\textwidth]{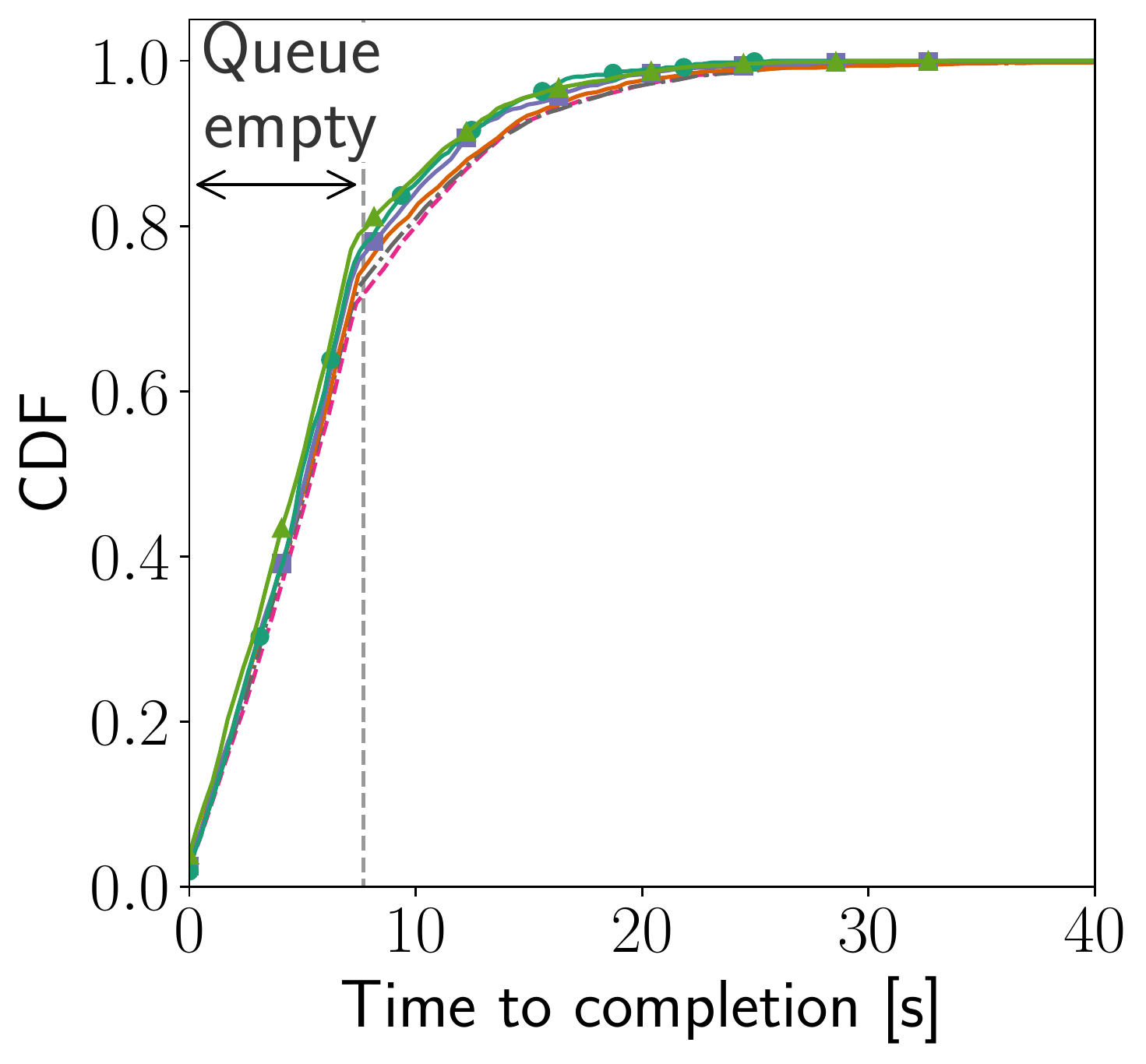}\label{fig:cfp_20}}
    \subfloat[CAP, TX interval=5s]{\includegraphics[width=.2425\textwidth]{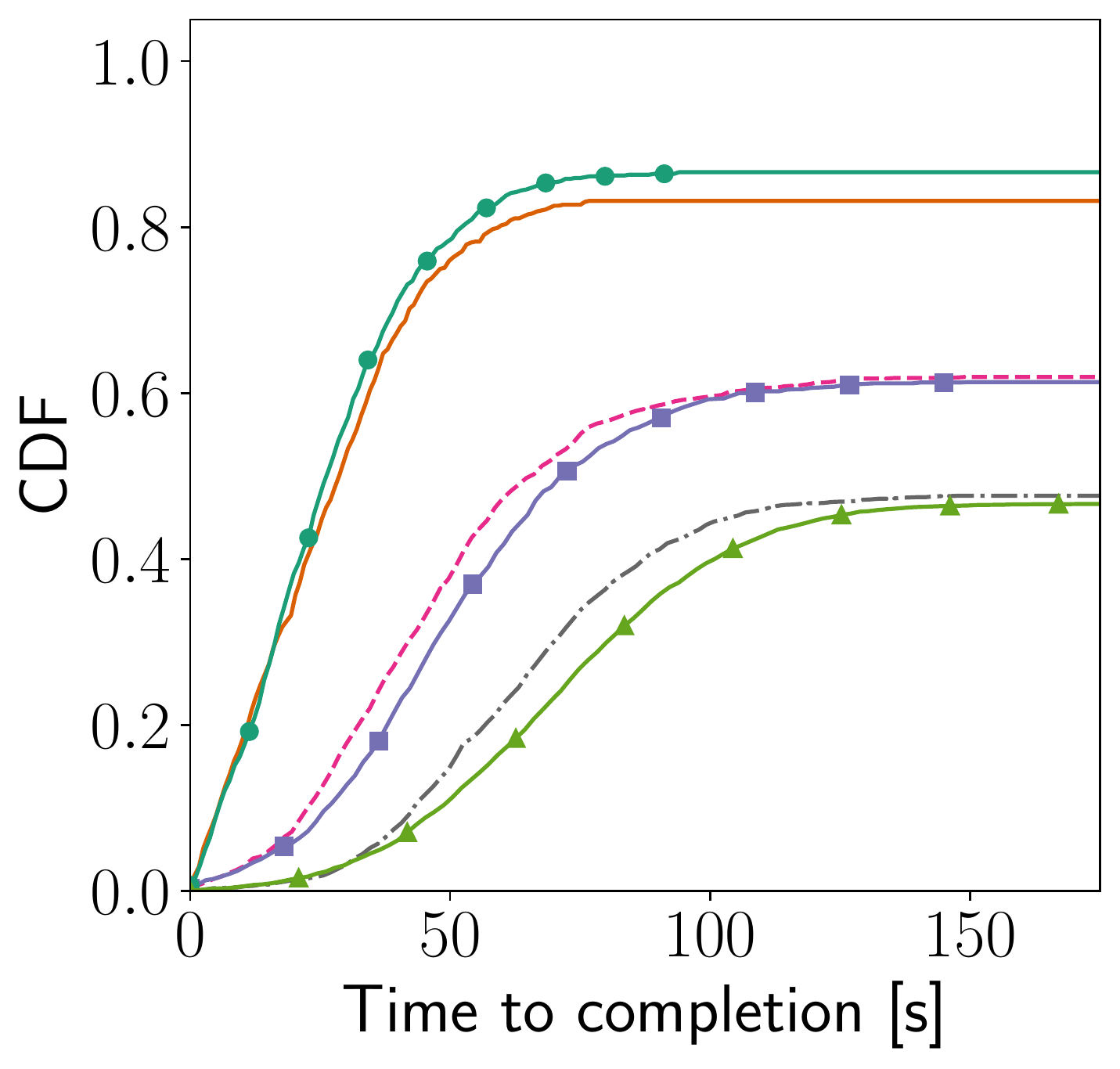}\label{fig:cap_5}}
    \subfloat[CFP, TX interval=5s]{\includegraphics[width=.2425\textwidth]{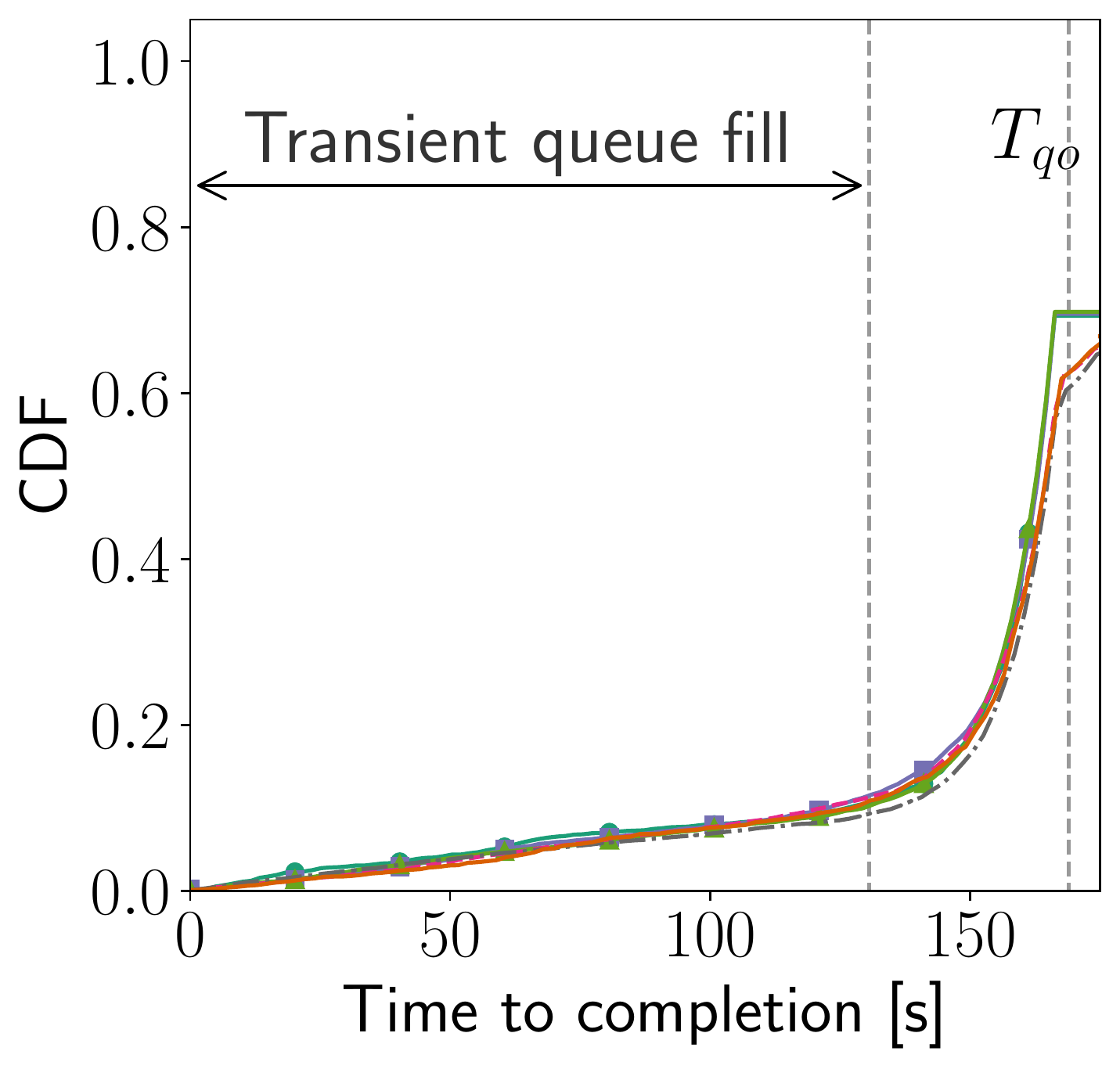}\label{fig:cfp_5}}
    \caption{Distribution of time to completion and PRR for varying number of nodes.}
    \label{fig:ttc_exponential_4}
\end{figure*}

Fig.~\ref{fig:ttc_exponential_4} shows the distribution of completion times and the packet reception ratio (PRR) for relaxed (Fig.~\ref{fig:cap_20}\&\ref{fig:cfp_20}) and stressed (Fig.~\ref{fig:cap_5}\&\ref{fig:cfp_5}) data transmissions. We separate CAP and CFP transmissions and show real-world measurements next so comparable simulation results for three network sizes (5/10/15 nodes).

\paragraph{Transmission during CAP}\label{sec:eval_cap} 
Our results show that the time to completion increases with an increase of network size (Fig.~\ref{fig:cap_20}) and a decrease of the transmission interval (Fig.~\ref{fig:cap_5}). Both cases increase the amount of on-air traffic.
As a result, the number of failed CCA attempts increases and a higher number of frames delay until CCA reports a clear channel.
Packets stay in the MAC queue until a re-attempt succeeds and the MAC receives a valid ACK. Hence, the average time to completion increases.
Frames that face a channel busy have a higher chance of collision. Note that CCA is prone to inaccuracies, hence, the MAC transmits a fraction of packets when the channel is actually busy which leads to wireless interference.

In the relaxed scenario (Fig.~\ref{fig:cap_20}), CSMA-CA leads to a PRR of 100\,\%, regardless of the network size.
In contrast, in the stressed scenario (Fig.~\ref{fig:cap_5}) the PR decreases drastically with the number of nodes ($<$50\,\% with 15 nodes). This is due to exceeded CCA attempts which leads to discarded packets already on the sender, and remaining packet collisions on-air.

Our experimental measurements converge to simulation results. Differing channel models in the simulation environment and the physical channel upfront explain the small variations.

\paragraph{Transmission during CFP}\label{sec:eval_cfp} 
In the relaxed scenario (Fig.~\ref{fig:cfp_20}), the time to completion does not vary with the network size, because transmissions occur during a dedicated GTS.
Hence, all transmissions are collision free and have the same throughput (one packet per multisuperframe).
When the queue is empty, the MAC queue schedules packets in the next available slot. As a result, the time to completion is upper bounded by the duration of a multisuperframe.

Reducing the transmission interval reduces the chances of transmission when the MAC queue is empty, due to stress and queue saturation.
The time to completion in the stressed scenario (Fig.~\ref{fig:cfp_5}) increases $>$10x, because the transmission interval is shorter than the duration of a multisuperframe. This puts the MAC queue is an unstable condition in which the MAC schedules packets when the queue is (nearly) overflowed.
In these cases, the MAC layer queueing delays transmissions by multiples of a multisuperframe duration.
Naturally, the fraction of packets with a time to completion upper bounded by one multisuperframe duration decreases.

The first transmissions in Fig.~\ref{fig:cfp_5} (0--60\,s) reflect an initial transient phase that fills the MAC queue sequentially. These packets have a short time to completion.
$T_{qo}$ represents the completion time of the last packet in a saturated queue, assuming the MAC transmits all queued packets without retransmissions.
Nonetheless, a small fraction of packets are still retransmitted and the time to completion exceeds $T_{qo}$.

The PRR does not vary with the network size, as a result of collision free transmissions. Hence, the cause of packet loss is either \one channel interference or \two MAC queue overflow.
In the relaxed scenario (Fig.~\ref{fig:cfp_20}), \one holds and reveals a high reception ratio of 99.8\,\%, due to the robust LoRa modulation.
The stressed scenario (Fig.~\ref{fig:cfp_5}), in contrast, shows the effect of \two and reveals a low reception rate of $\approx$ 65\,\% due to discarded packets by sender when the MAC queue is full.
Note that the PRR does not vary with the network size, even under stress.

Similarly to~Fig.~\ref{sec:eval_cap}, differences between simulation results and real world experiments are small and explained by a varying number of retransmissions.
While our simulations get along without retransmissions, the real world experiments incorporate sporadic retransmissions that increase the completion time marginally.

\section{Conclusion and Outlook}\label{sec:conclusion}


In this paper, we started from the observation that sound protocol analysis
needs complementary methods. To close the gap for DSME over LoRa, a
cutting-edge MAC proposal for decentralized, long-range IoT communication,
we contributed the first implementation of DSME-LoRa running on real
hardware. In contrast to existing simulations, we implemented DSME-LoRa
based on RIOT, an open-source IoT operating system. We evaluated the
performance on resource constrained hardware in an open-access testbed.
Our measurement results are on par with prior simulations.

Our experiments show that DSME-LoRa enables node-to-node communication in
long-range network scenarios.
Data transmission during the contention access period is subject to packet
collisions, which leads to degradation of time to completion and packet
reception ratio, when traffic increases. Nevertheless, the CAP is typically only used to negotiate CFP resources.
Data transmission during CFP, however, enables reliable packet delivery with deterministic latencies due to exclusive time-frequency slots.
The delivery ratio degrades with stress in the MAC queue, however, the network size does not affect the time to completion using CFP.


Our research agenda is twofold. First, we will explore low power
capabilities of DSME-LoRa on hardware. Second, we will adopt concepts
proposed by the IETF 6TiSCH group, taking advantage of built-in scheduling
features of DSME to enable multi-hop communication. Thereby, we will enable
IPv6 connectivity through the network subsystem of RIOT.


\bibliographystyle{IEEEtran}
\bibliography{own, ids, iot, layer2}

\end{document}